\documentclass[3p,times,twocolumn]{elsarticle}

\usepackage{ecrc}

\volume{00}

\firstpage{1}

\journalname{Nuclear Physics B Proceedings Supplement}

\runauth{P. Mosteiro et al.}

\jid{nuphbp}

\jnltitlelogo{Nuclear Physics B Proceedings Supplement}

\usepackage{amssymb}

\usepackage[figuresright]{rotating}

\newcommand{\Rb}{$^{87}$Rb}
\newcommand{\be}{\begin{equation}}
\newcommand{\ee}{\end{equation}}
\newcommand{\Be}{$^7$Be}
\newcommand{\pp}{$pp$}
\newcommand{\pep}{$pep$}
\newcommand{\B}{$^8$B}
\newcommand{\C}{$^{14}$C}
\newcommand{\Bi}{$^{210}$Bi}
\newcommand{\Po}{$^{210}$Po}
\newcommand{\Kr}{$^{85}$Kr}
\newcommand{\Pb}{$^{214}$Pb}
\newcommand{\K}{$^{40}$K}

\begin{document}

\begin{frontmatter}


\title{Low-energy (anti)neutrino physics with Borexino: \\
Neutrinos from the primary proton-proton fusion process in the Sun
}

\dochead{}


\author[l]{P.~Mosteiro\fnref{nowatrome}}
\author[h]{G.~Bellini}
\author[k]{J.~Benziger}
\author[s]{D.~Bick}
\author[e]{G.~Bonfini}
\author[q]{D.~Bravo}
\author[h]{B.~Caccianiga}
\author[p]{L.~Cadonati\fnref{nowatgt}}
\author[l,gssi]{F.~Calaprice}
\author[c]{A.~Caminata}
\author[e]{P.~Cavalcante}
\author[l]{\'A.~Chavarr\'ia}
\author[r]{A.~Chepurnov}
\author[h]{D.~D'Angelo}
\author[t]{S.~Davini}
\author[m]{A.~Derbin}
\author[t]{A.~Empl}
\author[g,mephi]{A.~Etenko}
\author[b,e]{K.~Fomenko}
\author[a]{D.~Franco}
\author[e]{F.~Gabriele}
\author[l]{C.~Galbiati}
\author[e]{S.~Gazzana}
\author[c]{C.~Ghiano}
\author[h]{M.~Giammarchi}
\author[n]{M.~G\"oger-Neff}
\author[l]{A.~Goretti}
\author[r]{M.~Gromov}
\author[s]{C.~Hagner}
\author[t]{E.~Hungerford}
\author[e]{Al.~Ianni}
\author[l]{An.~Ianni}
\author[f]{V.~Kobychev}
\author[b]{D.~Korabl\"ev}
\author[t]{G.~Korga}
\author[a]{D.~Kryn}
\author[e]{M.~Laubenstein}
\author[dresden]{B.~Lehnert}
\author[n]{T.~Lewke}
\author[g,mephi]{E.~Litvinovich}
\author[e]{F.~Lombardi}
\author[h]{P.~Lombardi}
\author[h]{L.~Ludhova}
\author[g]{G.~Lukyanchenko}
\author[g,mephi]{I.~Machulin}
\author[q]{S.~Manecki}
\author[i]{W.~Maneschg}
\author[c,gssi]{S.~Marcocci}
\author[n]{Q.~Meindl}
\author[h]{E.~Meroni}
\author[s]{M.~Meyer}
\author[h]{L.~Miramonti}
\author[d]{M.~Misiaszek}
\author[e]{M.~Montuschi}
\author[m]{V.~Muratova}
\author[n]{L.~Oberauer}
\author[a]{M.~Obolensky}
\author[j]{F.~Ortica}
\author[p]{K.~Otis}
\author[c]{M.~Pallavicini}
\author[q,n]{L.~Papp}
\author[c]{L.~Perasso}
\author[p]{A.~Pocar}
\author[h]{G.~Ranucci}
\author[e]{A.~Razeto}
\author[h]{A.~Re}
\author[j]{A.~Romani}
\author[e]{N.~Rossi}
\author[l]{R.~Saldanha}
\author[c]{C.~Salvo}
\author[n]{S.~Sch\"onert}
\author[i]{H.~Simgen}
\author[g,mephi]{M.~Skorokhvatov}
\author[b]{O.~Smirnov}
\author[b]{A.~Sotnikov}
\author[g]{S.~Sukhotin}
\author[u,g,e]{Y.~Suvorov}
\author[e]{R.~Tartaglia}
\author[c]{G.~Testera}
\author[a]{D.~Vignaud}
\author[q]{R.~B.~Vogelaar}
\author[n]{F.~von Feilitzsch}
\author[u]{H.~Wang}
\author[mainz]{J.~Winter}
\author[d]{M.~Wojcik}
\author[l]{A.~Wright}
\author[mainz]{M.~Wurm}
\author[b]{O.~Zaimidoroga}
\author[c]{S.~Zavatarelli}
\author[dresden]{K.~Zuber}
\author[d]{G.~Zuzel}

\address[l]{Physics Department, Princeton University, Princeton, NJ 08544, USA}
\address[b]{Joint Institute for Nuclear Research, Dubna 141980, Russia}
\address[h]{Dipartimento di Fisica, Universit\`a degli Studi e INFN, Milano 20133, Italy}
\address[p]{Amherst Center for Fundamental Interactions and Physics Department, University of Massachusetts, Amherst, MA 01003, USA}
\address[a]{APC, Univ. Paris Diderot, CNRS/IN2P3, CEA/Irfu, Obs. de Paris, Sorbonne Paris Cit\'e, France}
\address[c]{Dipartimento di Fisica, Universit\`a e INFN, Genova 16146, Italy}
\address[q]{Physics Department, Virginia Polytechnic Institute and State University, Blacksburg, VA 24061, USA}
\address[e]{INFN Laboratori Nazionali del Gran Sasso, Assergi 67010, Italy}
\address[dresden]{Department of Physics, Technische Universita?t Dresden, 01062 Dresden, Germany}
\address[mephi]{National Research Nuclear University MEPhI (Moscow Engineering Physics Institute), 115409 Moscow, Russia}
\address[g]{NRC Kurchatov Institute, Moscow 123182, Russia}
\address[d]{M. Smoluchowski Institute of Physics, Jagiellonian University, Krakow, 30059, Poland}
\address[u]{Physics and Astronomy Department, University of California Los Angeles (UCLA), Los Angeles, CA 90095, USA}
\address[s]{Institut f\"ur Experimentalphysik, Universit\"at Hamburg, Germany}
\address[n]{Physik Department, Technische Universit\"at M\"unchen, Garching 85747, Germany}
\address[i]{Max-Plank-Institut f\"ur Kernphysik, Heidelberg 69029, Germany}
\address[j]{Dipartimento di Chimica, Universit\`a e INFN, Perugia 06123, Italy}
\address[m]{St. Petersburg Nuclear Physics Institute, Gatchina 188350, Russia}
\address[t]{Department of Physics, University of Houston, Houston, TX 77204, USA}
\address[f]{Kiev Institute for Nuclear Research, Kiev 06380, Ukraine}
\address[r]{Lomonosov Moscow State University Skobeltsyn Institute of Nuclear Physics, Moscow 119234, Russia}
\address[k]{Chemical Engineering Department, Princeton University, Princeton, NJ 08544, USA}
\address[gssi]{Gran Sasso Science Institute (INFN), 67100 L'Aquila, Italy}
\address[mainz]{Institut f\"ur Physik, Johannes Gutenberg Universit\"at Mainz, 55122 Mainz, Germany}

\fntext[nowatrome]{Now at INFN, 00185 Roma, Italy}
\fntext[nowatgt]{Now at Georgia Tech, Atlanta, GA 30332, USA}



\begin{abstract}
The Sun is fueled by a series of nuclear reactions that produce the energy that makes it shine. 
The primary reaction is the fusion of two protons into a deuteron, a positron and a neutrino.
These neutrinos constitute the vast majority of neutrinos reaching Earth, providing us with key information about what goes on at the core of our star.
Several experiments have now confirmed the observation of neutrino oscillations by detecting neutrinos from secondary nuclear processes in the Sun; this is the first direct spectral measurement of the neutrinos from the keystone proton-proton fusion.
This observation is a crucial step towards the completion of the spectroscopy of $pp$-chain neutrinos, as well as further validation of the LMA-MSW model of neutrino oscillations.
\end{abstract}




\end{frontmatter}


\section{Solar neutrinos}
\label{sec:solar_neutrinos}
\begin{figure}
\begin{center}
\includegraphics[width=3in]{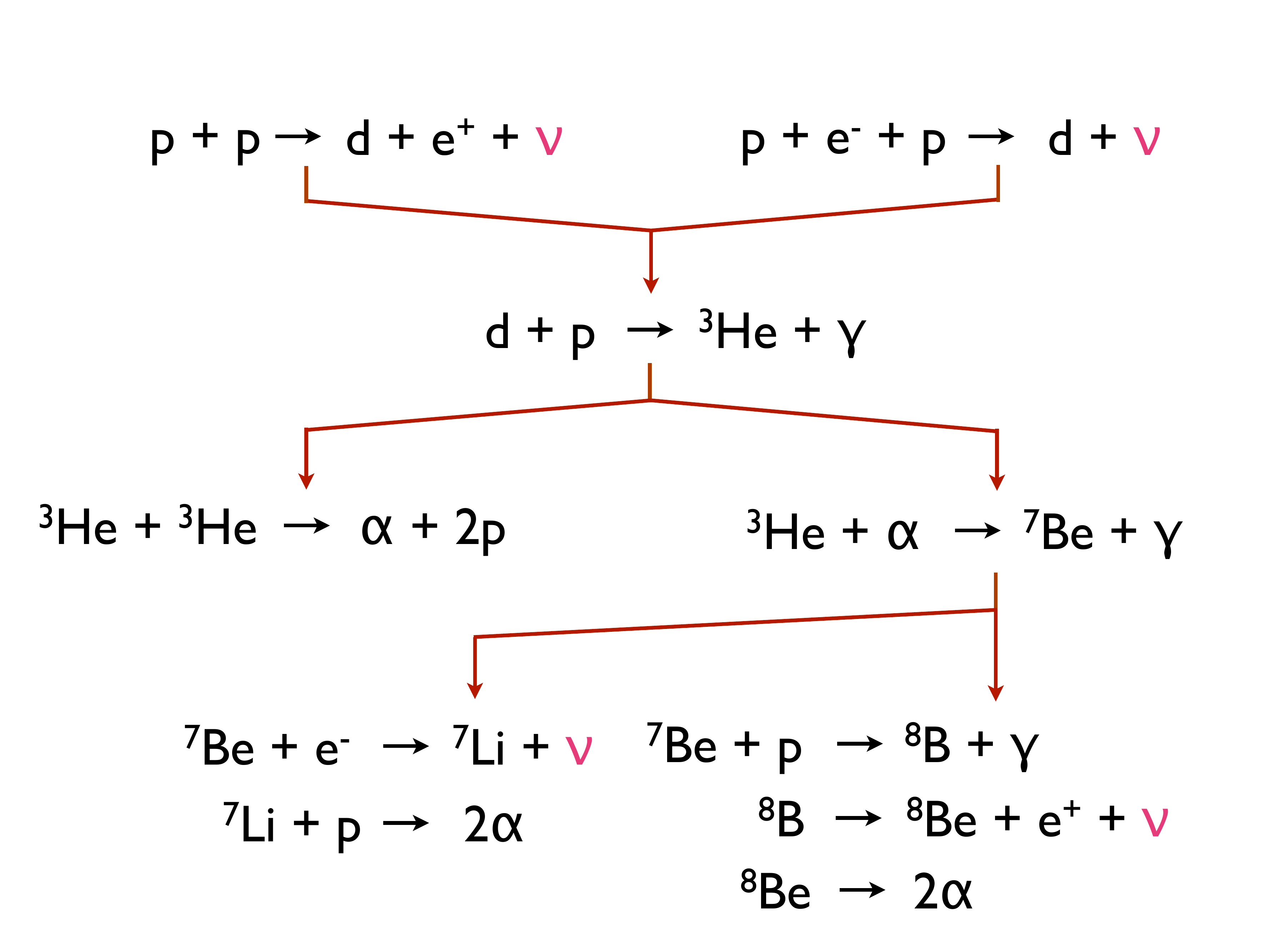}
\caption{Main nuclear reactions that make up the $pp$ chain~\cite{clayton}. The neutrinos produced by the chain are highlighted in magenta. Neutrinos are named after the parent particles that produce them (\pp, \pep, \Be, \B).
$pep$, $^7$Be, and $^8$B neutrinos have all been previously measured with Borexino~\cite{pep,7Be,B8}, leaving only the $pp$ neutrinos (top left) and the extremely faint $hep$ neutrinos~\cite{bahcall} (not shown here) undetected prior to this work.}
\label{fig:pp-chain}
\end{center}
\end{figure}

We know that 
the Sun
is fueled by nuclear reactions~\cite{bethe-energy-production}. In particular, the ``effective reaction'' that takes place is the conversion of hydrogen (p) into helium (He), 
\be
4{\rm p}\rightarrow {\rm ^{4}He}+2{\rm e}^++2\nu_{\rm e},
\ee
with a net release of kinetic energy and emission of electron neutrinos $\nu_{\rm e}$.
There are two main ways of converting protons to He nuclei that take place in stars: the $pp$ chain and the CNO cycle~\cite{clayton}. The contribution from each of these processes depends on the size, temperature and age of the star~\cite{bethe-energy-production}.
The main way of producing energy in the Sun and in stars of similar mass and age is the $pp$ chain~\cite{BahcallViewgraphs}.
Fig.~\ref{fig:pp-chain} shows the main reactions that are responsible for it.

The different neutrinos in the chain are often referred to by their parent particles. We thus speak of $pp$ neutrinos as those produced in the top left of the chain, whose parents are simply two protons.
Though most of the energy and neutrino flux in the Sun comes from the $pp$ chain, more massive and hotter stars produce much more significant amounts of energy through the CNO cycle~\cite{bethe-energy-production}. In addition, a small component of the Sun's neutrino flux also comes from the CNO cycle.

The challenge in studying neutrinos is intimately connected with the reason why they are of interest in the field of solar astrophysics: they can only interact via the weak force. Thus, they travel almost undisturbed from the core of the Sun to the surface of the Earth, providing us with key information about the inner workings of our star.

The energy spectrum of solar neutrinos 
as predicted by the Standard Solar Model~\cite{serenelli} (SSM)
is shown in~\cite{BahcallViewgraphs}.
The end-point energy is $E^{\rm max}_\nu=420$\,keV~\cite{bahcall}.
There are two currently accepted values for the total \pp-neutrino flux produced by the Sun, corresponding to the \emph{low-metallicity} and \emph{high-metallicity} models~\cite{serenelli}.
The values are $5.98\times(1\pm0.006)\times10^{10}$\,cm$^{-2}$\,s$^{-1}$ and $6.03\times(1\pm0.006)\times10^{10}$\,cm$^{-2}$\,s$^{-1}$, respectively.

Various measurements of solar neutrinos~\cite{pep,7Be,B8,davis,SNO} have proven that the Sun is fueled by nuclear reactions and that neutrinos undergo lepton flavor transformation~\cite{SNO,SuperK,MS} (also known as \emph{oscillations}), which itself means that at least some of the neutrinos must be massive.
It also implies that the flux quoted above will contain electron-, muon- and tau-neutrinos upon arrival at Earth.

Previous experiments~\cite{gallex1,gallex2,sage} were able to extract the detection rate of \pp\ neutrinos from an integral of the solar neutrino spectrum, by subtracting the rates of other solar neutrino components~\cite{SNO,7Be}.
No experiment has thus far been able to directly and spectrally detect \pp\ neutrinos, that is, to observe interactions of \pp\ neutrinos and measure their energies. The main obstacle has been the overwhelming background at the low energies characteristic of \pp\ neutrinos (see Sec.~\ref{sec:borexino}).

\section{Borexino}
\label{sec:borexino}
Borexino~\cite{alimontiDetector} is a liquid scintillator detector located at the Laboratori Nazionali del Gran Sasso, in central Italy. It detects neutrinos via their elastic scattering interactions with electrons in the scintillator.
Scintillation light is collected by \emph{photomultipliers} (PMTs) surrounding the \emph{inner detector}. Outside the inner detector, a water tank as a detector of Cherenkov light, tagging cosmogenic muons with very high efficiency~\cite{firstmuon}.

When combining the \pp-neutrino flux provided by the Standard Solar Model (Sec.~\ref{sec:solar_neutrinos}) with the latest values of the neutrino oscillation parameters~\cite{beringer,BahcallRadiative} and with information about the Borexino scintillator~\cite{SaldanhaThesis}, we can calculate the expected rate of \pp-induced electron recoils in Borexino to be $131\pm2$\,(d$\cdot$100\,t)$^{-1}$ (assuming the high-metallicity SSM; the low-metallicity number can be obtained by rescaling according to the fluxes).

Scintillation light induces signals in $\sim$500 photomultipliers (PMTs) per MeV of deposited energy~\cite{Cosmogenics}. This means that the threshold of the detector can be set as low as $\sim$50\,keV.
The end-point of the \pp-neutrino energy spectrum translates via relativistic kinematics to an end-point energy of the \pp-neutrino-induced electron recoil spectrum $E^{\rm max}_R=264$\,keV.
Thus, the low energy threshold is especially important for this analysis.

The \pp-neutrino-induced electron-recoil rate is obtained through a fit of the energy distribution of events against the known signal and background spectral shapes. 
Our energy estimator is the number of PMTs detecting photons within 230\,ns after the beginning of the physics event. This is converted to energy using energy-response and energy-resolution functions tuned in previous analyses and re-validated in the current study~\cite{long-paper,MosteiroThesis}.

To reduce radioactive background coming from external detector components, we only study events inside the central \emph{fiducial volume} (see Sec.~\ref{sec:pp}). 
The positions of events are determined by maximizing the likelihood of the observed PMT detection time distribution~\cite{galbiati-mccarty,ctf-background}. The performance of the algorithm has been tested with calibration sources~\cite{BxCalib}; further details can be found in~\cite{long-paper}.
At the energies relevant for the \pp\ analysis, the position reconstruction code has an uncertainty of 20\,cm.
This was estimated by comparing the reconstructed and nominal positions of radioactive sources placed inside Borexino during a calibration campaign~\cite{B8,BxCalib}.

The main backgrounds in this analysis are \C\, intrinsic to the organic liquid scintillator, and its pile-up. Other backgrounds include \Bi, \Po, \Kr\ and \Pb. The rate of \C\ $\beta$-decay in Borexino is $\sim40\,({\rm s}\cdot{\rm 100\,t})^{-1}$~\cite{SaldanhaThesis}, four orders of magnitude larger than the expected \pp-induced recoil rate (above).
The strategy for assessing the real background rates is discussed in Sec.~\ref{sec:pp}.

\section{\pp\ analysis}
\label{sec:pp}
Data for the present analysis was collected between January 2012 and June 2013, containing the first 408 live days of data of Borexino Phase-II. These data are characterized by reduced levels of some of the most prominent radiogenic backgrounds, most notably \Kr, \Bi\ and \Po, thanks to a series of successful purification campaigns~\cite{long-paper}.
Basic selection criteria have been described previously in~\cite{long-paper}, in the context of Borexino Phase-I.
In particular, the criteria in the current analysis are: \\
\textit{(1)} No coincidence with muon events; \\
\textit{(2)} Event triggered by a scintillation event in the inner detector (Sec.~\ref{sec:borexino}); \\
\textit{(3)} No electronic noise from PMTs or electronics racks; \\
\textit{(4)} No coincidence with isotopic decays in the radon branch of the uranium series; \\
\textit{(5)} Position reconstruction within the fiducial volume; for this analysis, the fiducial volume is a truncated sphere with $R=3.021$\,m, $|z|=1.67$\,m; the fiducial volume is 86\,m$^3$, or 75.5\,t.

\begin{figure}
\begin{center}
\includegraphics[width=3in]{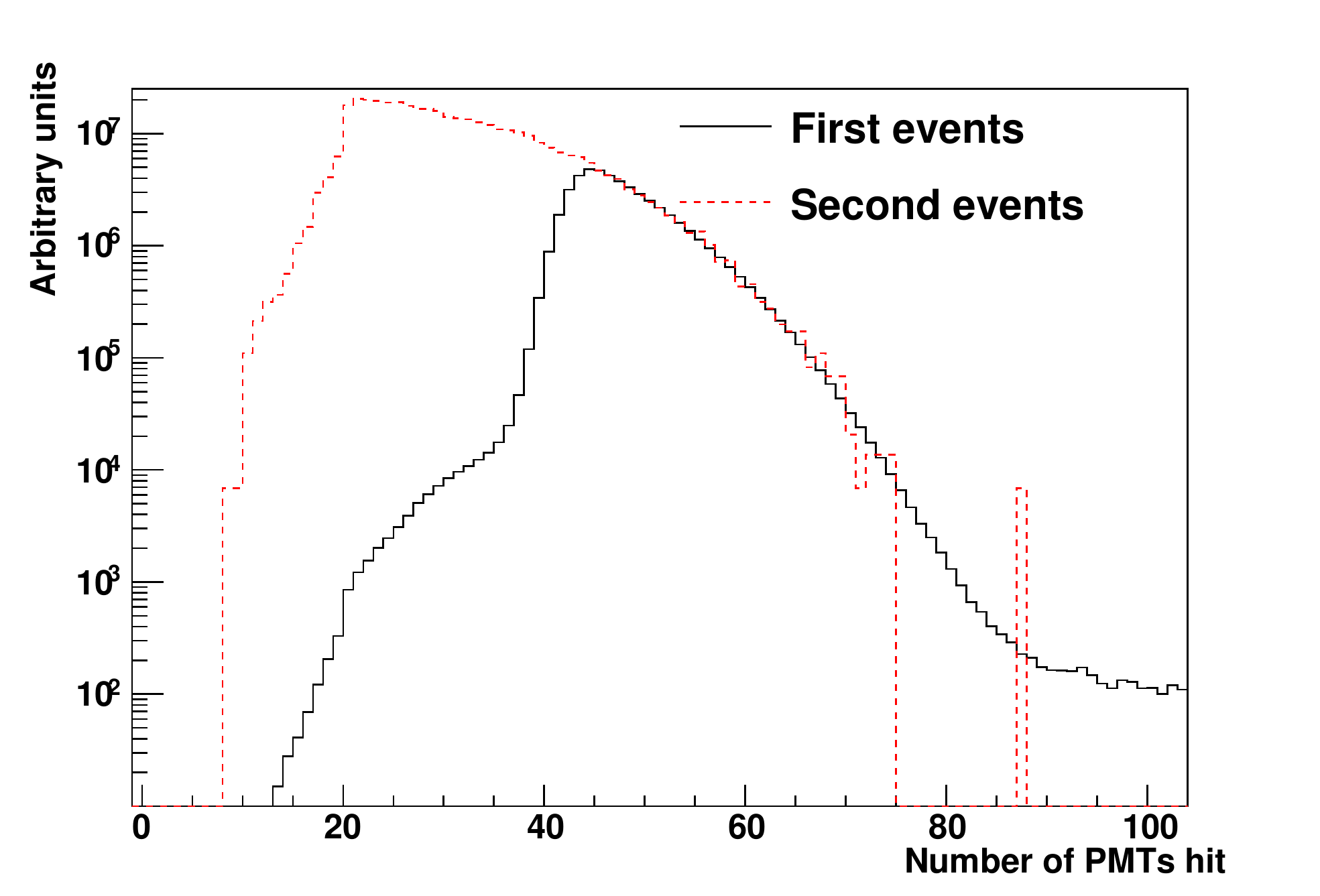}
\caption{Spectra of first and second events in trigger windows acquired during the live time considered in the \pp\ analysis, with the cuts outlined in the text. Second clusters are not subject to the trigger threshold characteristic of Borexino~\cite{long-paper}, and are thus useful for the independent estimate of the \C\ rate.}
\label{fig:second_cluster}
\end{center}
\end{figure}

As explained in Sec.~\ref{sec:borexino}, most of the remaining background is \C\ $\beta$-decays. To estimate its rate independently from the main analysis, we looked at a sample of data in which the event causing the trigger is followed by a second event within the same trigger window (16-$\mu$s long, while physical events last $\sim$100-500\,ns).
These second events are not subject to the trigger threshold, so they register down to much lower energies.
The spectra of first and second events are shown in Fig.~\ref{fig:second_cluster}.
By fitting this spectrum against the known \C\ spectral shape~\cite{kuzminov}, we obtain a \C\ rate in Borexino of 40$\pm$1\,Bq per 100\,t. The error accounts for statistical and systematical uncertainties, most notably due to uncertainty in the \C\ spectral shape~\cite{kuzminov,mortara,wietfeldt} and fit conditions.

Due to its high rate, \C\ naturally generates a considerable amount of pile-up: two physical events occur so close to each other in time that the detector fails to resolve them. With the above measurement of the \C\ rate, the expected rate of \C-\C\ pile-up is $\sim$100 per day per 100\,t. This is comparable to the expected \pp\ neutrino interaction rate (Sec.~\ref{sec:borexino}). In addition, the end-point of the \C-\C\ pile-up spectrum is 312\,keV (twice the \C\ spectrum end-point~\cite{toi}), also close to the end-point of the \pp-induced recoil spectrum, $E^{\rm max}_R$. This underlines the importance of a careful assessment of pile-up events, both \C-\C\ and other pile-up.

The pile-up component can be determined using an independent, data-driven method called \emph{synthetic pile-up}. Real triggered events with no cuts are artificially overlapped with random data obtained from the ends of real trigger windows, uncorrelated with the triggering event. 
By overlapping four random data samples with each physical trigger, we obtain a sample of synthetic events that has four times the exposure of the real data set used.
The synthetic events are then reconstructed using the same software used for real events, and selected with the same criteria. Using this method, we obtained the true rate and spectral shape of pile-up in our detector.
This method naturally accounts for the effect of fiducialization on pile-up events.

With the \C\ and pile-up rates fixed, we obtained the \pp\ rate by fitting the data spectrum in a window of 165-590\,keV against the known spectral shapes. The fit was performed with a tool named \emph{spectral-fitter}, previously used in \Be~\cite{7Be} and \pep~\cite{pep} analyses. The tool was improved to work on any Borexino analysis.
In particular, we included a treatment of the scintillator energy response to mono-energetic electrons at high statistics; a modified treatment of the energy resolution at low energy; and the introduction of synthetic pile-up.

In addition to \pp, \C\ and pile-up, other species included in the fit are other solar neutrinos (\Be, CNO and \pep), \Kr, \Bi, \Po\ and \Pb. The \Be\ rate was fixed to the rate previously obtained by Borexino~\cite{7Be}; CNO and \pep\ rates were fixed to the values predicted by the high-metallicity Standard Solar Model; remaining background rates are left free in the fit, except for \Pb, which is fixed to the value measured by looking at $^{214}$Bi-$^{214}$Po coincidences~\cite{long-paper}. The scintillator light yield and two energy resolution parameters are free to vary in the fit.

\begin{figure}
\begin{center}
\includegraphics[width=3in]{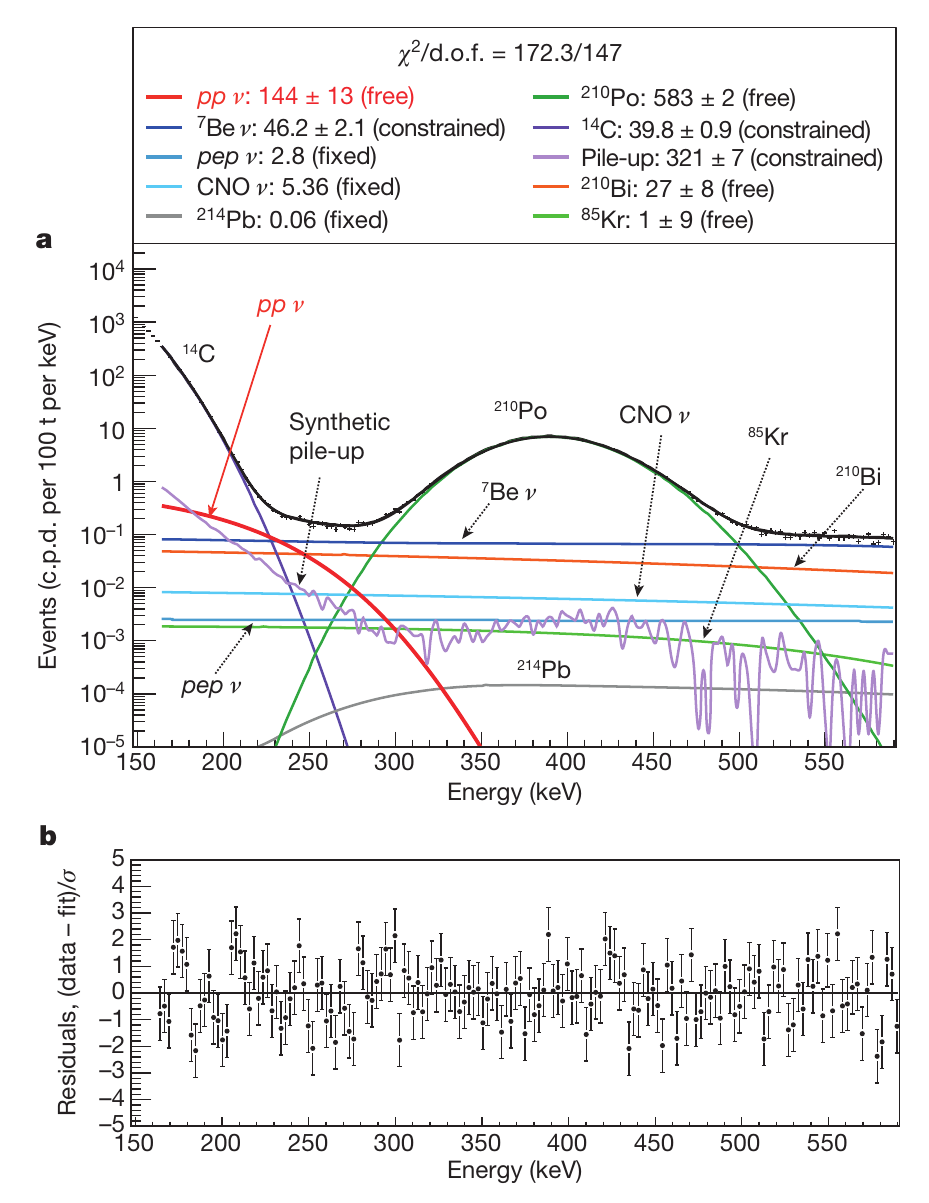}
\caption{\textbf{a.} Fit of the data in our energy region of interest to extract the value of the \pp\ neutrino interaction rate. The \Be\ rate displayed corresponds only to the higher-energy branch~\cite{7Be}. Rates are displayed in (d$\cdot$100\,t)$^{-1}$, except for that of \C, which is in (s$\cdot$100\,t)$^{-1}$.
Uncertainties are statistical only.
Species labelled ``constrained'' are free to vary, but a penalty is applied to the likelihood if the value chosen by the fitter differs from the expected value~\cite{MosteiroThesis}.
\textbf{b.} Residuals of the fit. Systematic effects are discussed in the text.}
\label{fig:pp_fit}
\end{center}
\end{figure}

The energy spectrum after cuts, together with the best fit, is shown on Fig.~\ref{fig:pp_fit}.
The uncertainties shown in the figure are statistical only.

\begin{figure}
\begin{center}
\includegraphics[width=3in]{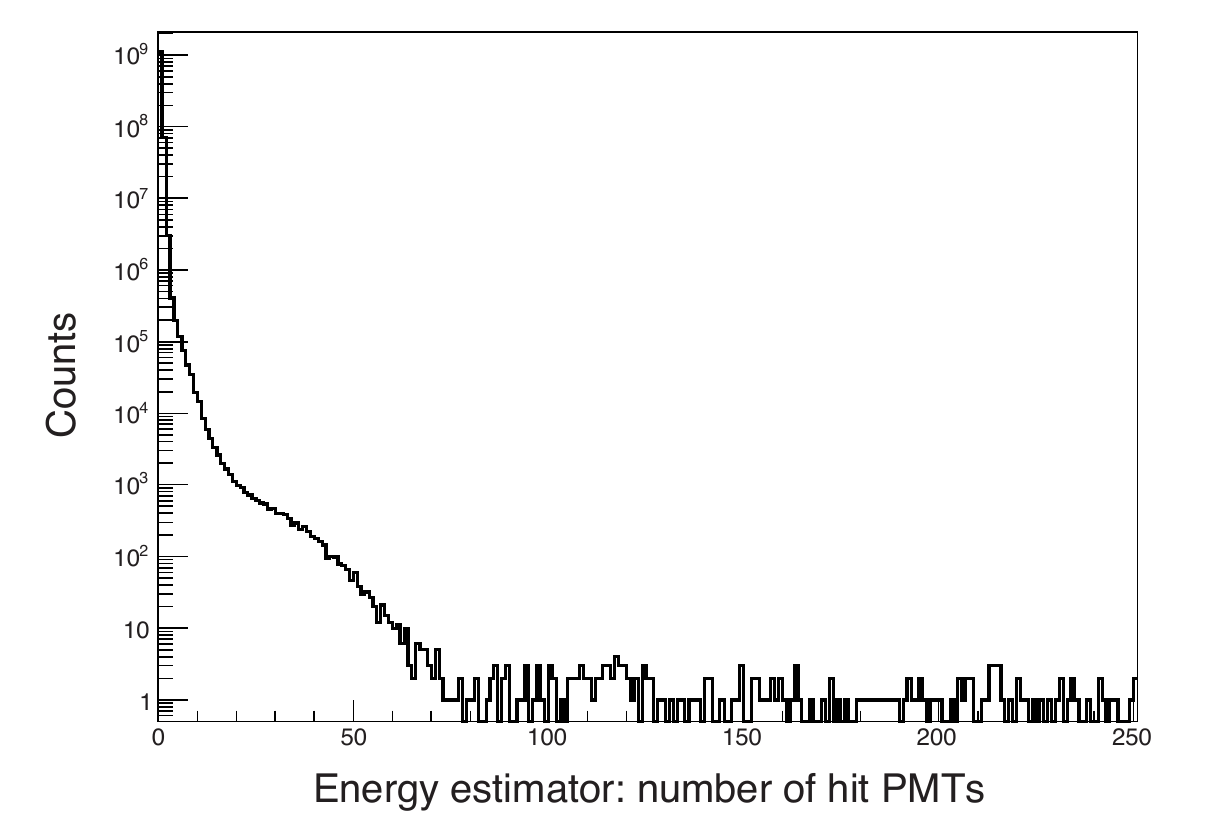}
\caption{Spectrum obtained by slicing random triggers into smaller time windows with the same duration as the windows used in regular data for the computation of our energy estimator, as explained in Sec.~\ref{sec:borexino}. This is the spectrum obtained by Borexino without a trigger threshold. It is used in the current analysis for an alternative estimate of the pile-up rate.}
\label{fig:convolution}
\end{center}
\end{figure}

We considered an alternative method to account for pile-up, known as \emph{convolution method}.
Random 16-$\mu$s-long samples of data are regularly collected by Borexino~\cite{long-paper}.
These data are sliced into time windows of the same duration as the window used for our energy estimator (Sec.~\ref{sec:borexino}).
By computing the energy estimator inside each of these small windows, we generate the distribution shown in Fig.~\ref{fig:convolution}.
This distribution is the spectrum collected by Borexino without a trigger threshold, and it includes electronic noise, as well as all physical events generated by signal and background species. Pile-up can be generated by convolving each spectral component (neutrinos and backgrounds) with this spectrum.

\begin{figure}
\begin{center}
\includegraphics[width=3in]{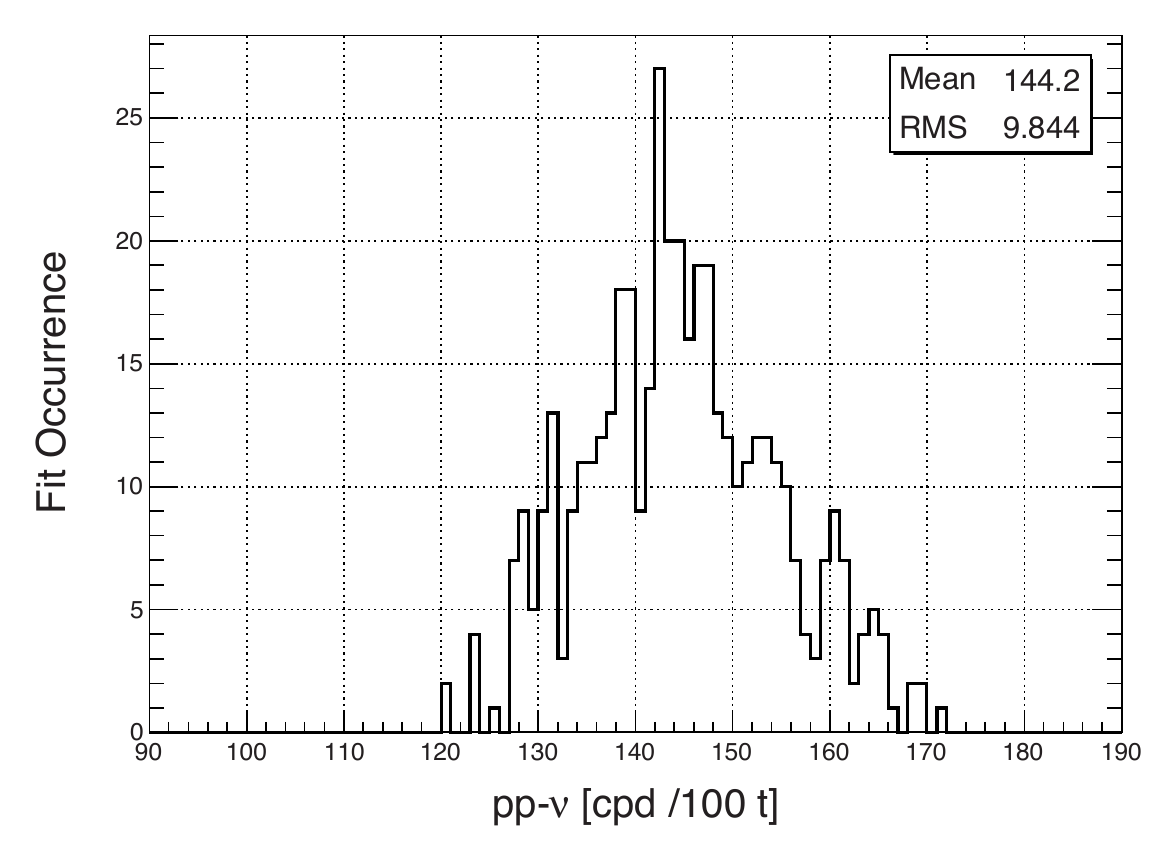}
\caption{\pp\ neutrino interaction rate as obtained from many fits where fit conditions are varied to account for the systematic effects outlined in the text. The mean value and root mean square of this distribution are our central value and systematic uncertainty, respectively.}
\label{fig:fit_histo}
\end{center}
\end{figure}

Fit conditions were varied to estimate the systematic uncertainty. Effects considered include variations in: \\
\textit{(1)} The pile-up evaluation method (\emph{synthetic} and \emph{convolution}); \\
\textit{(2)} The length of the time window in which we count PMTs hit (Sec.~\ref{sec:borexino}); \\
\textit{(3)} The definition of the fiducial volume; \\
\textit{(4)} The expected value of the CNO neutrino detection rate; \\
\textit{(5)} The fit start and end points. \\
Fits were performed using many different combinations of all these effects. The histogram of the rates obtained in all these conditions is shown in Fig.~\ref{fig:fit_histo}.
The root mean square of the distribution is 9.8\,(d$\cdot$100\,t)$^{-1}$.
In addition, the resolution of the position reconstruction algorithm (20\,cm; Sec.~\ref{sec:borexino}) induces a 2\% uncertainty in the determination of the fiducial volume. 
These uncertainties are added in quadrature to obtain the total systematic uncertainty of our measurement.
The final \pp\ neutrino interaction rate is~\cite{pp}
\be
R_{pp} = 144 \pm 13\,{\rm (stat)} \pm 10.\,{\rm (syst)}\,{\rm (d\cdot100\,t)}^{-1}
\ee
We also studied possible variations in the result caused by: \\
\textit{(1)} Uncertainties in the spectral shapes of \C~\cite{kuzminov,mortara,wietfeldt} and \Bi~\cite{long-paper,daniel,flothmann}; \\
\textit{(2)} Variations of the expected \pep\ rate within the measured and theoretical uncertainties; \\
\textit{(3)} Leaving the \C\ rate free in the fit; \\
\textit{(4)} Varying the veto time after muons cross the detector. \\
No significant variations were observed.

We also studied as a potential source of background the $\beta$-emitter \Rb, with a Q-value of 283.3\,keV~\cite{toi}, very similar to the end-point of the \pp-induced electron recoil spectrum $E^{\rm max}_R$ (Sec.~\ref{sec:borexino}). Using the known crustal abundances~\cite{webelements}, isotopic abundances and half-lives~\cite{toi} of \Rb\ and \K, we can calculate the ratio between their natural activies to be 1 to 4, respectively.
Under the assumption that the relative abundances of \Rb\ and \K\ in Borexino are equal to their relative crustal abundances, and knowing that the rate of \K\ in Borexino is $<0.4$\,(d$\cdot$100\,t)$^{-1}$~\cite{pep}, we can constrain the \Rb\ rate to $<$0.1\,(d$\cdot$100\,t)$^{-1}$.

\section{Interpretation}
\label{sec:interp}
The \emph{electron neutrino survival probability}, $P_{ee}$, is the probability for a solar neutrino produced as electron-type to arrive at Earth as an electron-neutrino. It can be computed using the relations in~\cite{deholanda,nakamura}, the most up-to-date oscillation parameters and interaction cross-sections from~\cite{beringer,BahcallRadiative,fogli} and the electron number density in Borexino, (3.307$\pm$0.003)$\times10^{31}$\,(100\,t)$^{-1}$~\cite{7Be}.
Using that number and our measured \pp\ neutrino interaction rate, we can calculate the \pp\ neutrino flux to be $(6.6\pm0.7)\times10^{10}$\,cm$^{-2}$\,s$^{-1}$, consistent with the SSM prediction (see Sec.~\ref{sec:solar_neutrinos}).
The uncertainty of the present measurement does not allow us to distinguish between the high- and low-metallicity solar models.

\begin{figure}
\begin{center}
\includegraphics[width=3in]{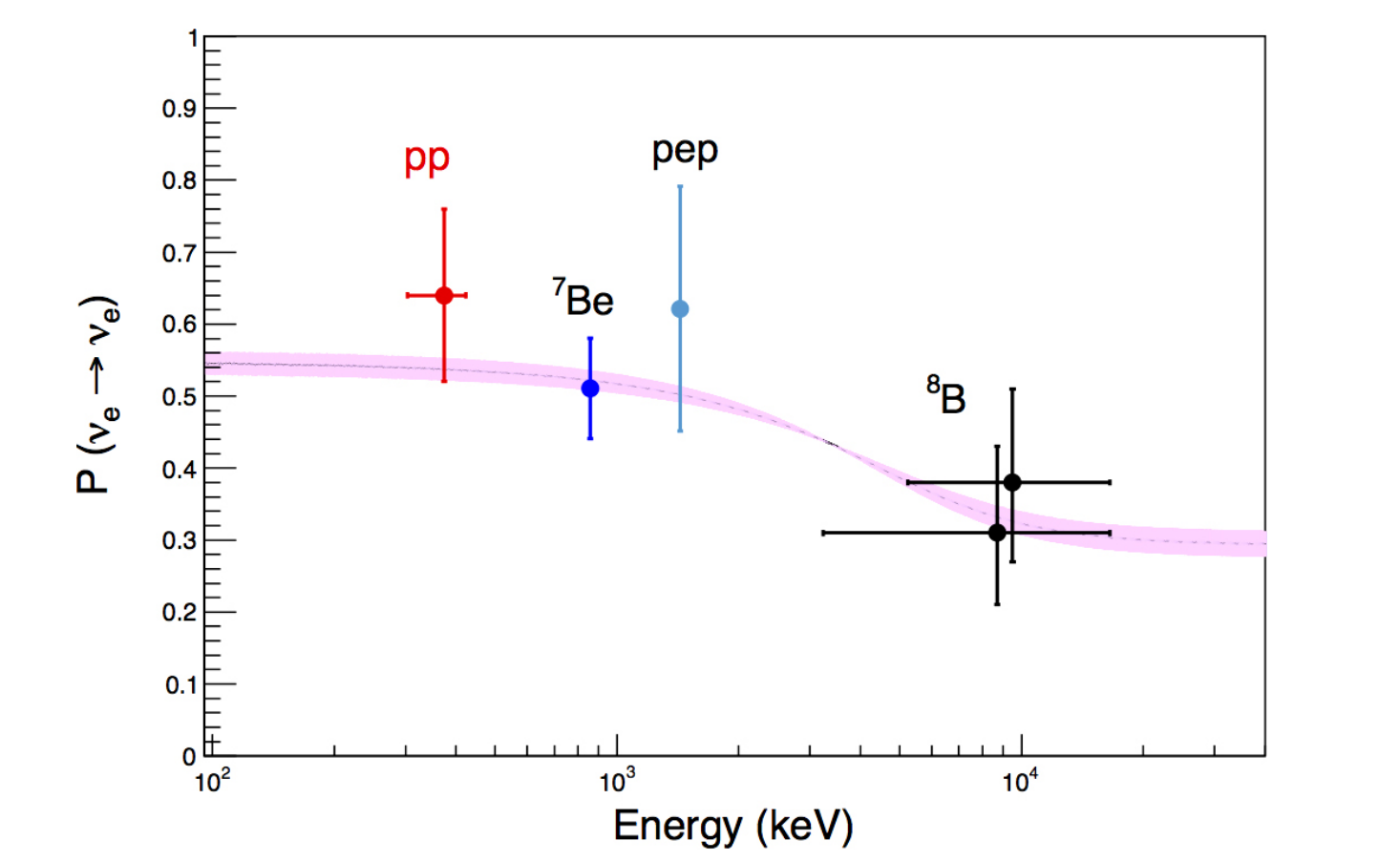}
\caption{Survival probability $P_{ee}$ for all solar neutrino species measured by Borexino. The red marker is the present result; \Be\ is from~\cite{7Be}; \pep, from~\cite{pep}; and \B, from~\cite{B8}. \Be\ and \pep\ are monoenergetic neutrinos. \B\ and \pp\ are continuous spectra, and the reported energies correspond to the energy ranges studied in the analyses. 
The vertical error bars are 1$\sigma$ uncertainties.
The curve is the theoretical prediction from the Large Mixing Angle (LMA) solution of the Mikheyev-Smirnov-Wolfenstein (MSW) effect~\cite{beringer}; the violet band is the 1$\sigma$ uncertainty. The curve was calculated for \B\ neutrinos, as they are the only species that covers the entire energy range shown and their production region in the Sun includes all the other species as well.
}
\label{fig:pee}
\end{center}
\end{figure}

Conversely, if we use the \pp\ neutrino flux predicted by the SSM as an input, we can calculate the survival probability $P_{ee}=0.64\pm0.12$, consistent with expectation from the LMA-MSW solution~\cite{BahcallLMASMALOW}, as shown in Fig.~\ref{fig:pee}.

Finally, the fact that the present measurement of the \pp\ neutrino flux is consistent with expectation justifies the assumption made by the SSM that the photon and neutrino luminosities from the Sun can be correlated; \textit{i.e.}, that the Sun has been stable for the past 10$^5$ years. This invalidates a previous hypothesis that changes in the solar core over these time scales could be responsible for terrestrial global epochs~\cite{variability}.




\nocite{*}
\bibliographystyle{elsarticle-num}
\bibliography{borexino-pp}







\end{document}